\documentclass[twocolumn,amsmath,amssymb,aps,prb,floatfix,superscriptaddress]{revtex4-2}

\usepackage[utf8]{inputenc}

\usepackage{epsfig}

\usepackage{graphicx}

\usepackage{multirow}

\usepackage{color}

\usepackage[colorlinks=true,allcolors=blue]{hyperref}

\usepackage{orcidlink}

\usepackage{latexsym}

\usepackage[Large,FIGTOPCAP]{subfigure}

\usepackage{IEEEtrantools}

\usepackage{blkarray}

\usepackage{mathtools}

\begin{document}

\title{Resolving the Magnetic Ground State and Field-Induced Transitions in magnetic Dirac semimetal candidate $\text{EuMnSb}_{2}$}

\author{Yiu-Fung Chiu,\orcidlink{0000-0001-9673-6167}}
\email{yiufung.chiu@physics.ox.ac.uk}

\affiliation{Department of Physics, University of Oxford, Clarendon Laboratory, Oxford OX1 3PU, United Kingdom}

\affiliation{Diamond Light Source, Harwell Campus, Didcot OX11 0DE, United Kingdom}

\author{Jian-Rui Soh,\orcidlink{0000-0003-2542-6086}}
\affiliation{Quantum Innovation Centre (Q.InC), Agency for Science Technology and Research (A*STAR),
2 Fusionopolis Way, Singapore 138634, Singapore}
\affiliation{Centre for Quantum Technologies, National University of Singapore, 3 Science Drive 2, Singapore 117543, Singapore}
\author{Sanjay Sharma}
\affiliation{Department of Physics, University of Warwick, Coventry, CV4 7AL, United Kingdom}

\author{J. Alberto Rodríguez-Velamazán,\orcidlink{0000-0002-8505-5232}}
\affiliation{Institut Laue-Langevin, 71 avenue des Martyrs, CS 20156, 38042 Grenoble Cedex 9, France}

\author{John Singleton,\orcidlink{0000-0001-6109-6905}}
\affiliation{Department of Physics, University of Oxford, Clarendon Laboratory, Oxford OX1 3PU, United Kingdom}
\affiliation{National High Magnetic Field Laboratory Pulsed-Field Facility, TA-35, MS-E536, Los Alamos National Laboratory, Los Alamos, New Mexico 87545, USA}

\author{Eugen Weschke,\orcidlink{0000-0002-2141-0944}}
\affiliation{Helmholtz-Zentrum Berlin für Materialien und Energie, BESSY II, Albert-Einstein-Straße 15, 12489 Berlin, Germany}

\author{Oleksandr Prokhnenko,\orcidlink{0000-0002-5376-1765}}
\affiliation{Helmholtz-Zentrum Berlin für Materialien und Energie, BESSY II, Albert-Einstein-Straße 15, 12489 Berlin, Germany}

\author{Oksana Zaharko,\orcidlink{0000-0001-5521-3124}}
\affiliation{PSI Center for Neutron and Muon Sciences, Forschungsstrasse 111, 5232 Villigen, PSI, Switzerland}

\author{Dharmalingam Prabhakaran,\orcidlink{0000-0002-7769-9716}}

\affiliation{Department of Physics, University of Oxford, Clarendon Laboratory, Oxford OX1 3PU, United Kingdom}

\author{Stephen J. Blundell,\orcidlink{0000-0002-3426-0834}}

\affiliation{Department of Physics, University of Oxford, Clarendon Laboratory, Oxford OX1 3PU, United Kingdom}

\author{Paul A. Goddard,\orcidlink{0000-0002-0666-5236
}}
\affiliation{Department of Physics, University of Warwick, Coventry, CV4 7AL, United Kingdom}

\author{Andrew~T.~Boothroyd,\orcidlink{0000-0002-3575-7471}}

\email{andrew.boothroyd@physics.ox.ac.uk}

\affiliation{Department of Physics, University of Oxford, Clarendon Laboratory, Oxford OX1 3PU, United Kingdom}

\date{\today}

\begin{abstract}

The magnetic structure of a magnetic topological semimetal EuMnSb$_{2}$ is investigated in fields up to 30\,T  using polarized and unpolarized neutron diffraction, pulsed-field x-ray magnetic circular dichroism and pulsed-field magnetometry. We determine the zero-field magnetic structures of the Eu and Mn sublattices, and find that magnetic transitions induced by applied fields below 2\,T correspond to changes in the magnetic order of the Eu spins alone  without detectable perturbation to the order of the Mn spins. An additional magnetic transition is observed at fields close to the saturation field for the Eu spins.
We present a mean-field model which  describes key features of the magnetic behavior and allows us to estimate the dominant Eu--Eu and Eu--Mn exchange interactions responsible for the coupling between magnetism and electronic topology. 

\end{abstract}

\maketitle

\section{Introduction}

Magnetic topological semimetals have attracted interest because of the possibility of controlling topological electronic states through their coupling to magnetism and applied magnetic fields~\cite{Bernevig2022,Li2020,Zou2019}. The ability to manipulate electronic topology and the associated spin-polarized surface states has potential for applications in spintronics~\cite{Tokura,Bernevig,Wang,Zou,Xu}.

A case of particular interest is EuMnSb$_{2}$, which shows a range of interesting magnetic, magnetotransport and topological behaviors~\cite{Yi,Zhu,Soh,Gong,Sun,Zhang21,Zhang,Wilde,Zhao,Yin}. Electronic structure calculations indicate that  EuMnSb$_{2}$ is either a gapped Dirac semimetal or a Weyl semimetal, depending on the assumed magnetic structure~\cite{Sun,Wilde}. Angle-resolved photoemission spectroscopy (ARPES) measurements showed that EuMnSb$_{2}$ possesses a Dirac-like linear dispersion of the valence bands near $E_\textrm{F}$~\cite{Soh}, and quantum oscillations measurements revealed a Berry phase with a non-trivial temperature dependence~\cite{Zhang21,Zhang,Zhao}. Magnetotransport data show that magnetic ordering on the Eu magnetic sublattice strongly influences the electrical conductivity~\cite{Yi, Zhu,Soh,Sun,Zhang,Yin}, and the coupling between large spin-only magnetic moments on the Eu and Mn atoms is expected to drive metamagnetic phases.  Moreover, the metamagnetic phases in EuMnSb$_{2}$ could provide opportunities to field-tune the topological properties. Thus, a detailed determination of the intrinsic magnetic structures and magnetic interactions in EuMnSb$_{2}$ is essential to understand the interplay between electronic topology and magnetism.

The crystal structure and transport properties reported for EuMnSb$_2$ depend sensitively on the growth method and resulting composition. Crystals grown using Sn flux are generally orthorhombic (space group \textit{Pnma}) and exhibit semiconducting transport~\cite{Yi,Soh,Yin}. Representative room-temperature lattice parameters for Sn flux crystals are $a=22.57$\,\AA, $b=4.38$\,\AA, and $c=4.41$\,\AA~\cite{Soh}. 
In contrast, Zhang \textit{et al.}~obtained metallic, tetragonal  EuMn${_{0.95}}$Sb$_2$ ($P4/nmm$) using a stoichiometric self-flux method~\cite{Zhang}. The floating-zone-grown, Mn-rich EuMn${_{1.1}}$Sb$_2$ sample reported by Gong \textit{et al.}~was orthorhombic and showed weakly insulating transport with similar crystal lattice parameters as samples grown with Sn-flux~\cite{Gong,Yin}, whereas crystals grown with excess Sb flux have also been reported to be orthorhombic and metallic~\cite{Wilde,Yin}. Yin \textit{et al.}~associated a small difference in the $a$-axis lattice parameter with the two types of crystal (semiconducting and metallic), although variations in chemical composition or defect concentration introduced during growth may also contribute~\cite{Yin}. These differences indicate that the structural, magnetic, and electronic properties depend not only on the nominal growth route but also on small deviations from stoichiometry and possible flux incorporation.

\begin{figure*}
    \centering
    \includegraphics[width=\textwidth]{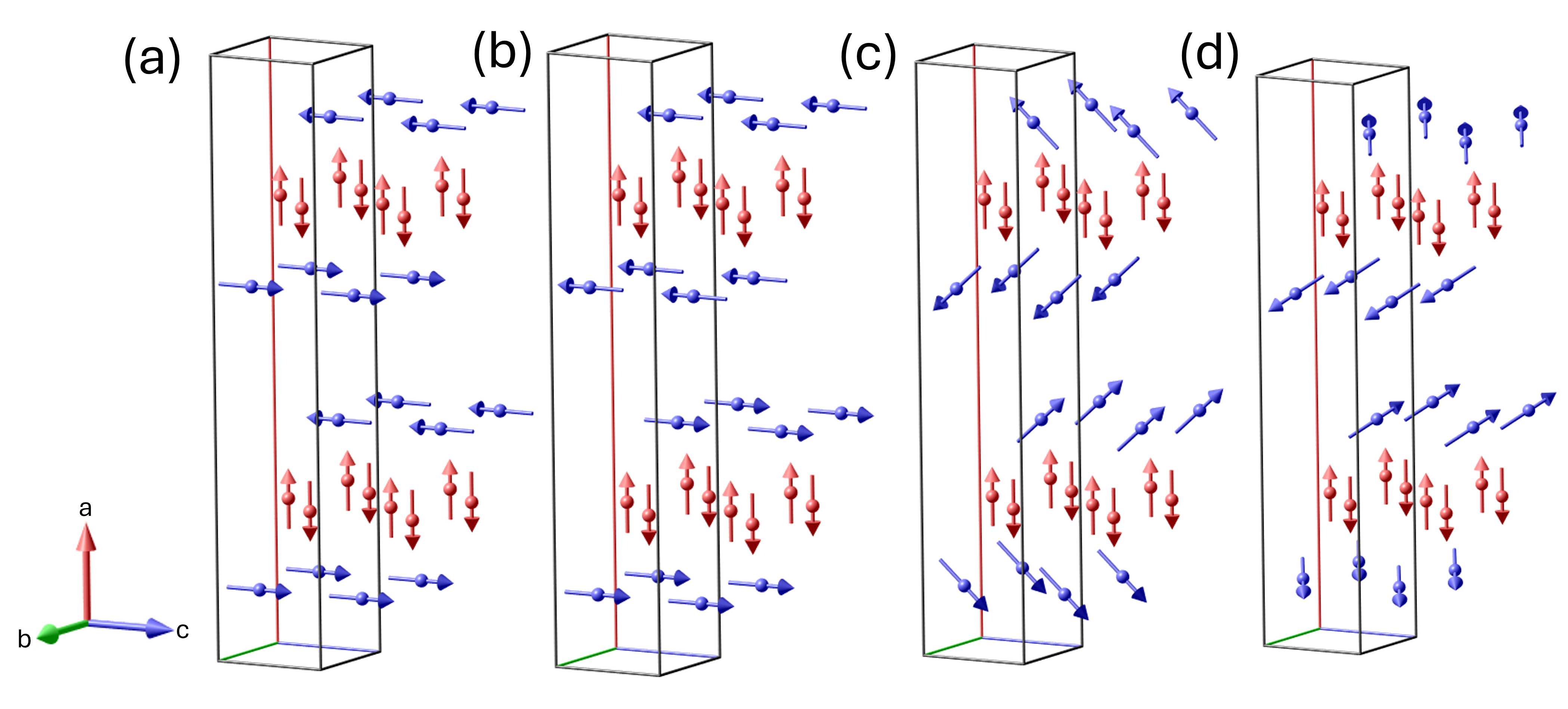}
    \caption{Magnetic structures of EuMnSb$_{2}$ at base temperature from neutron diffraction studies on different samples. Mn moments are red and Eu are blue. (a) Powder sample obtained by crushing Sn-flux-grown crystals~\cite{Soh}. (b)  Tetragonal EuMn$_{0.95}$Sb$_{2}$ single crystal grown by a stoichiometric self-flux method~\cite{Zhang}. (c) Orthorhombic EuMn$_{1.1}$Sb$_{2}$ single crystal grown by a floating-zone method~\cite{Gong}. (d) Orthorhombic EuMnSb$_2$  single crystal grown using excess Sb flux~\cite{Wilde}.}
    \label{fig:1}
\end{figure*}

Three magnetic transition temperatures have been reported in EuMnSb$_2$ in zero applied field~\cite{Yi,Zhu,Soh,Gong,Sun,Zhang21,Zhang,Wilde,Zhao,Yin}: $T_{\text{Mn}} = 320-350$\,K, corresponding to antiferromagnetic (AFM) ordering of the Mn sublattice, and $T_{\text{Eu1}} \simeq$ 23\,K and $T_{\text{Eu2}} \simeq$ 9\,K, corresponding to the onset of AFM order of the Eu spins ($T_\textrm{Eu1}$) followed by a reorientation of the Eu spins ($T_\textrm{Eu2}$). No fewer than four previous neutron diffraction studies of the zero-field magnetic structures have been reported~\cite{Soh,Zhang,Gong,Wilde}. While all agree on the structure adopted by the Mn spins below  $T_{\text{Mn}}$ --- a C-type collinear AFM with spins pointing along the $a$ axis, see Fig.~\ref{fig:1} --- reports differ on the Eu spin structures. Figure~\ref{fig:1}(a) shows an A-type AFM structure with stacking $(+-+-)$  along $a$ and Eu spins pointing along $c$ reported in the powder diffraction study of Soh \textit{et al.}~\cite{Soh}, while Fig.~\ref{fig:1}(b) shows the same A-type AFM except with stacking $(++--)$ along $a$ from the  measurements by Zhang \textit{et al.} on single crystals~\cite{Zhang}. Figures~\ref{fig:1}(c)--(d) depict the results of two other single-crystal studies. Gong \textit{et al.} \cite{Gong} found
a noncollinear canted A-type AFM with the spins lying in the $ac$ plane at an angle of $41^\circ$ from the $a$ axis,
Fig.~\ref{fig:1}(c), and Wilde \textit{et al.}~\cite{Wilde} found two magnetic phases: for $T_{\text{Eu2}} < T < T_{\text{Eu1}}$ that shown in Fig.~\ref{fig:1}(c),
and for $T < T_{\text{Eu2}}$ the non-collinear AFM with spin components along $a$, $b$ and $c$, Fig.~\ref{fig:1}(d).

Several factors may explain these discrepancies. First, the very strong nuclear absorption cross-section of natural Eu ($\sigma_\textrm{a} = 4530$\,b at a neutron wavelength of 1.8\,\AA) means that severe corrections depending on the shape of the sample must be applied to the measured Bragg peak intensities. Second, partial twinning in the $bc$ plane also complicates the analysis of single-crystal diffraction intensities. Third, as mentioned above, there are differences in the physical properties of single crystals of EuMnSb$_2$ grown by different methods~\cite{Yin}.

In this work, we aim to provide a conclusive determination of the magnetic structures in semiconducting EuMnSb$_{2}$ by using the technique of spherical neutron polarimetry (SNP), which does not suffer from the neutron self-absorption problem because it depends only on ratios of intensities measured with opposite neutron polarization, whereupon the absorption factor cancels. 
Our results in zero field are broadly consistent with the spin structures reported by Wilde \textit{et al.}~\cite{Wilde}. We also provide X-ray magnetic circular dichroism (XMCD) data which evidence that the Mn spins remain unchanged throughout the investigated temperature and magnetic field range.  Finally, we report magnetization measurements in steady and  pulsed fields, and develop a mean-field model to describe the key magnetic interactions.  We used a partially twinned single crystal of EuMnSb$_{2}$ grown in Sn flux, and performed all measurements on the same crystal for consistency.

\section{Experimental details}

Bulk single crystals of EuMnSb$_{2}$ were grown by the Sn flux method described in Ref.~\onlinecite{Yi}. Crystals grown this way are reported to have the 1:1:2 element ratio to within 2\%~\cite{Yi}. X-ray diffraction measurements on the crystal used in this study indicated the ratio of $bc$ twins to be around 2:1, on a high resolution single crystal diffractometer (Rigaku Smartlab).  The crystalline quality was checked by Laue x-ray diffraction and found to be very high, with sharp Bragg peaks and no detectable secondary grains. The lack of any superconducting transition at low temperatures in electrical transport and magnetic measurements (see below) indicates that the crystal contains no significant Sn flux inclusions.

AC electrical transport data were performed in zero magnetic field $H$ using a Physical Properties Measurement System (PPMS, Quantum Design). A constant current of 10--100\,$\mu$A was applied parallel to $c$ and longitudinal voltage recorded as a function of temperature $T$ down to 1.8\,K. Low-field magnetic susceptibility $\chi(T)$ and magnetization $M(H)$ measurements were performed at temperatures down to 2\,K and in fields up to $\mu_0 H = 7$\,T with a Magnetic Properties Measurement System (MPMS, Quantum Design). Isothermal pulsed-field $M(H)$ measurements up to 35\,T with a typical pulse length of $\approx7$\,ms were performed at the Nicholas Kurti High Magnetic Field Laboratory, University of Oxford. Crystals were fixed within a PCTFE ampoule to prevent sample movement. The ampoule can be moved in and out of a $1500$-turn, $1.5$\,mm bore, $1.5$\,mm-long compensated-coil susceptometer constructed from $50$-gauge high-purity copper wire. When the sample is in the coil, the voltage induced in the coil is proportional to the rate of change in $M$ over time. The signal is integrated and the background data, measured with an empty coil under the same conditions, is subtracted to obtain $M(H)$. The magnetic field value is measured using a coaxial $10$-turn coil. A $^{3}$He cryostat provides temperature control and is used to attain temperatures down to $500$\,mK. The $M(H)$ measured in pulsed fields were calibrated against low-field data measured in the MPMS.

SNP was performed with the CryoPad device installed on the D3 diffractometer at the Institut Laue--Langevin, Grenoble \cite{Tasset1999,Lelievre2005}. A polarized beam of neutrons was obtained by Bragg diffraction from a crystal of ferromagnetic Hesuler alloy (Cu$_2$MnAl) oriented to diffract neutrons of wavelength $\lambda = 0.83$\,\AA. An erbium filter was
placed in the incident beam to suppress second-order contamination.  Nutator and precession fields controlled the direction of neutron polarization, and a helium-3 spin filter was used to analyze the polarization of the diffracted beam.  The quantity measured in SNP is the polarization matrix, defined by \cite{Boothroyd_book}
\begin{align}
    P_{\alpha\beta} = \frac{N_{\alpha\beta} - N_{\alpha\overline{\beta}}}{N_{\alpha\beta} + N_{\alpha\overline{\beta}}},
\end{align}
where $N_{\alpha\beta}$ and $N_{\alpha\overline{\beta}}$ are the number of Bragg-diffracted counts when the incident neutron polarization is along $\alpha$ and the scattered polarization is measured parallel and antiparallel to $\beta$, respectively, with $\alpha$, $\beta$ being along $x$, $y$, $z$. Here, $x$ is defined to be parallel to the scattering vector $\textbf{Q}$, $z$ is perpendicular to the scattering plane, and $y$ completes the right-handed set of Cartesian axes. The crystal was aligned with the $c$ axis of the majority twin vertical ($\parallel z$). Polarization matrices were measured for several $(H,K,0)$ peaks, where $H$ and $K$ are integer Miller indices, including purely magnetic peaks from the Eu spin structure, purely Mn magnetic peaks, and peaks with nuclear--magnetic interference. Standard corrections for the time decay of the filter efficiency were applied on the basis of measurements in the spin-flip channel at the ($010$) magnetic reflection, which is structurally forbidden. Measurements were made at temperatures of 2, 7.5 and 30\,K, and the Mag2Pol software \cite{Mag2Pol} was used to refine the magnetic structures.

Unpolarized neutron diffraction measurements were performed on the ZEBRA diffractometer at the SINQ (PSI), Switzerland, and on the D9 diffractometer at the Institut Laue Langevin, France. 
To distinguish how spin structures on the Eu and Mn sublattices evolve as a function of field, pulsed-field XMCD measurements were performed at the Eu $M_{5}$ and Mn $L_{3}$ edges, on the UE46\_PGM-1 beamline at the BESSY II light source, Berlin. Peak fields of 30\,T were achieved with a pulse length of 2\,ms. XMCD data were obtained from the difference of the total-electron yield absorption signals, for $\sigma +$ and $\sigma -$ polarised x-rays. Data were taken at a base temperature of 10\,K.

\section{Results}

\begin{figure}
    \centering
    \includegraphics[width=0.7 \columnwidth]{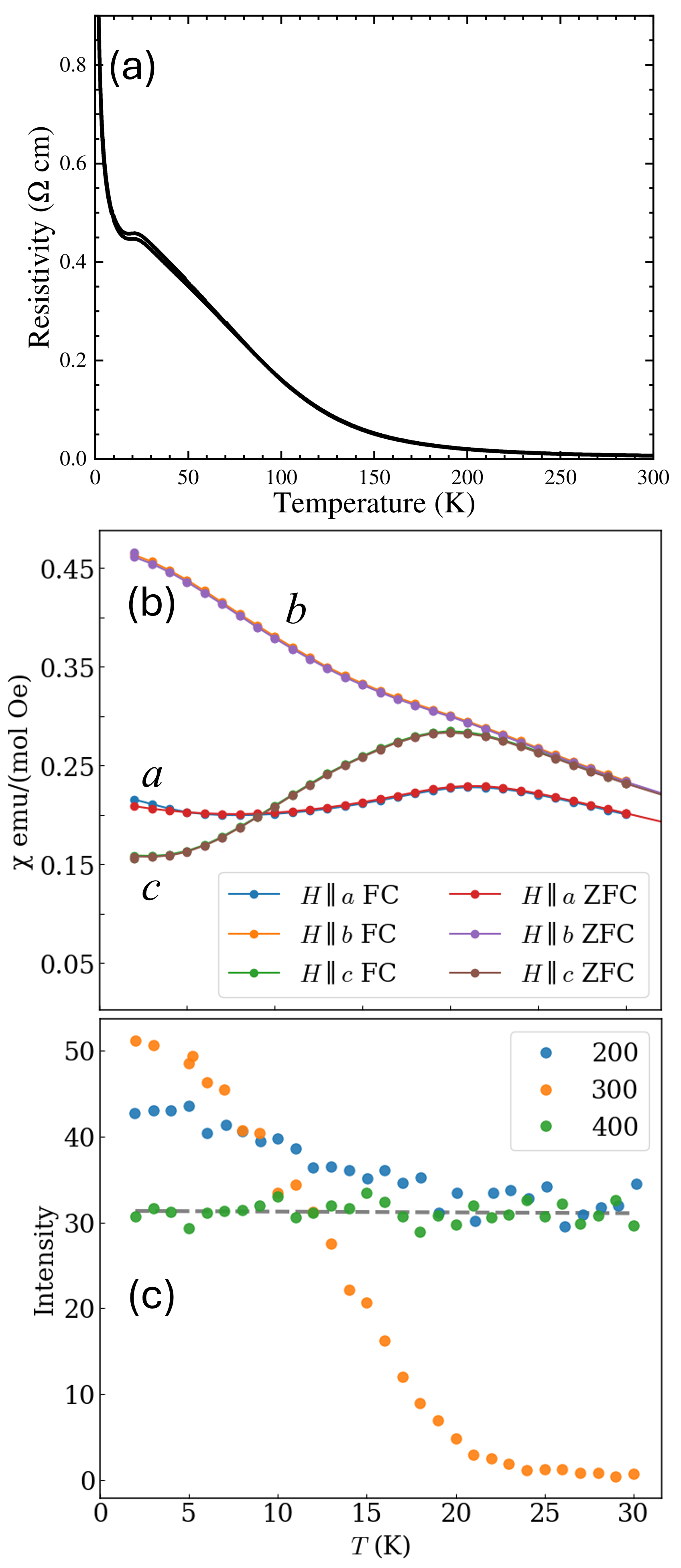}
    \caption{(a) Temperature dependence of the zero-field resistivity measured with current applied along the $c$ axis, showing semiconducting behaviour. (b) Magnetic susceptibility measurement in a field of 1000\,Oe applied parallel to the $a$, $b$ and $c$ axes. The Eu magnetic ordering transition is most clearly observed as a broad peak in the $a$- and $c$-axis curves near $T_{\textrm{Eu}1} \simeq 21$\,K. The $b$-axis curve exhibits a subtle upturn below $T_{\textrm{Eu}2} \simeq 9$\,K. (Data shown for $H \parallel b,c$-axis has been detwinned according to the 2:1 ratio.) (c) Unpolarized neutron diffraction data showing the onset of Eu magnetic order as signalled by the growth in intensity of the 300 reflection which is forbidden by the crystal symmetry and the Mn magnetic spin group symmetry. The 400 reflection has no magnetic contribution, while the 200 reflection has both structural and magnetic contributions. The grey line shows a linear fit of the 400 reflection.}
    \label{fig:4}
\end{figure}

Figure ~\ref{fig:4}(a) shows the zero-field $c$-axis resistivity of our Sn-flux-grown crystal. The resistivity increases strongly with decreasing temperature, confirming that the sample belongs to the semiconducting rather than metallic class of EuMnSb$_{2}$ crystals~\cite{Yi,Yin}. A shoulder in the temperature dependence is observed near $T_{\text{Eu1}} \simeq$ 21 K, consistent with the coupling between charge transport and the successive magnetic transitions of the Eu sublattice. This overall behaviour agrees with previous transport measurements on Sn-flux-grown EuMnSb$_{2}$~\cite{Yi,Yin}. 

The temperature dependence of the magnetic susceptibility in the range from 2 to 30\,K measured along the $a$, $b$ and $c$ axes is shown in Fig.~\ref{fig:4}(b). There is a maximum at $T\simeq 21$\,K in the $a$- and $c$-axis curves, with the peak in the latter being the more prominent, while the $b$-axis curve rises monotonically with decreasing temperature. This behavior is consistent with the transition to antiferromagnetic order at $T_{\text{Eu1}}$ reported previously, with the Eu spins aligned in the $ac$ plane. Below about 10\,K, the susceptibility has a slight upturn in the $a$ and $c$ directions with a corresponding reduction in the rate of increase along the $b$ direction. These  changes are likely associated with $T_{\text{Eu2}}$, and are consistent with a canting of the moments towards the $b$ axis.

Figure~\ref{fig:4}(c) displays unpolarized neutron diffraction over the same temperature range as Fig.~\ref{fig:4}(b).   The appearance of intensity at the (300) reflection below $T_{\text{Eu1}}$  is direct evidence for AFM order of the Eu spins, as this reflection is forbidden by both the crystallographic space group of EuMnSb$_2$ and the Mn AFM structure.  Moreover, we can deduce that there is a significant component of the ordered spins in the $bc$ plane since magnetic reflections of the type ($H00$) probe spin components perpendicular to $a$.  There is also a small increase in the (200) reflection below $T_{\text{Eu1}}$, but not in the (400) reflection. This indicates that there are both nuclear and magnetic contributions to (200) but only nuclear scattering at (400).

\begin{figure}
    \centering
    \includegraphics[width=0.8 \columnwidth]{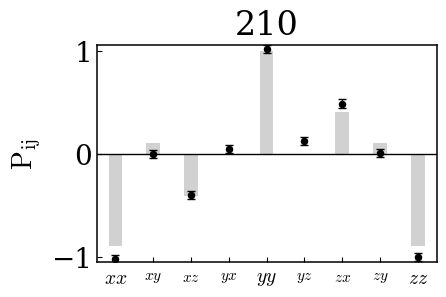}
    \caption{Polarisation matrix for 210 reflection measured at 30\,K, where only the Mn sublattice is ordered magnetically. Grey bars are calculated from the Mn magnetic structure shown in Fig.~\ref{fig:1} with zero ordered moment on Eu. }
    \label{fig:210 30K}
\end{figure}

We now present our SNP measurements of the spin structures in EuMnSb$_2$. To confirm the previously reported magnetic structure of the Mn spins, we performed SNP measurements at 30K on the  $(210)$ and (2$-$10) Bragg peaks. These peaks are equivalent by symmetry and include contributions from both nuclear and magnetic scattering. The measured polarization matrix $P_{ij}$ for $(210)$ is represented in Fig.~\ref{fig:210 30K}. The pure nuclear and pure magnetic scattering is contained in the diagonal elements $P_{xx}$, $P_{yy}$ and $P_{zz}$. Off-diagonal terms in $P_{ij}$ arise from nuclear--magnetic interference scattering and scattering from certain non-collinear spin components. To within error, $P_{xy} = P_{yx} = 0$, which is consistent with no $c$ axis component of the Mn spins. On the other hand, $P_{xz} = -P_{zx} \ne 0$ means that there is nuclear--magnetic interference scattering from the spin component in the $ab$ plane. As shown in Fig.~\ref{fig:210 30K}, the polarization matrix calculated for a C-type AFM structure with Mn spins of magnitude $\sim$4.2 $\mu_{\textrm B}$ parallel to $a$ agrees well with the data, consistent with previous reports~\cite{Gong,Zhang,Wilde}. We note that the existence of non-zero off-diagonal matrix elements implies a magnetic domain imbalance. In fact, our refinements show that the magnetic phase is almost entirely formed from one of the two time-reversal domains of the Mn AFM structure. It is not clear why a single magnetic domain formed given that the two time-reversal domains should have the same energy. Since its origin is unclear and not central to the magnetic structure determination, we leave this as an open question.

\begin{figure}
    \centering
    \includegraphics[width=0.8 \columnwidth]{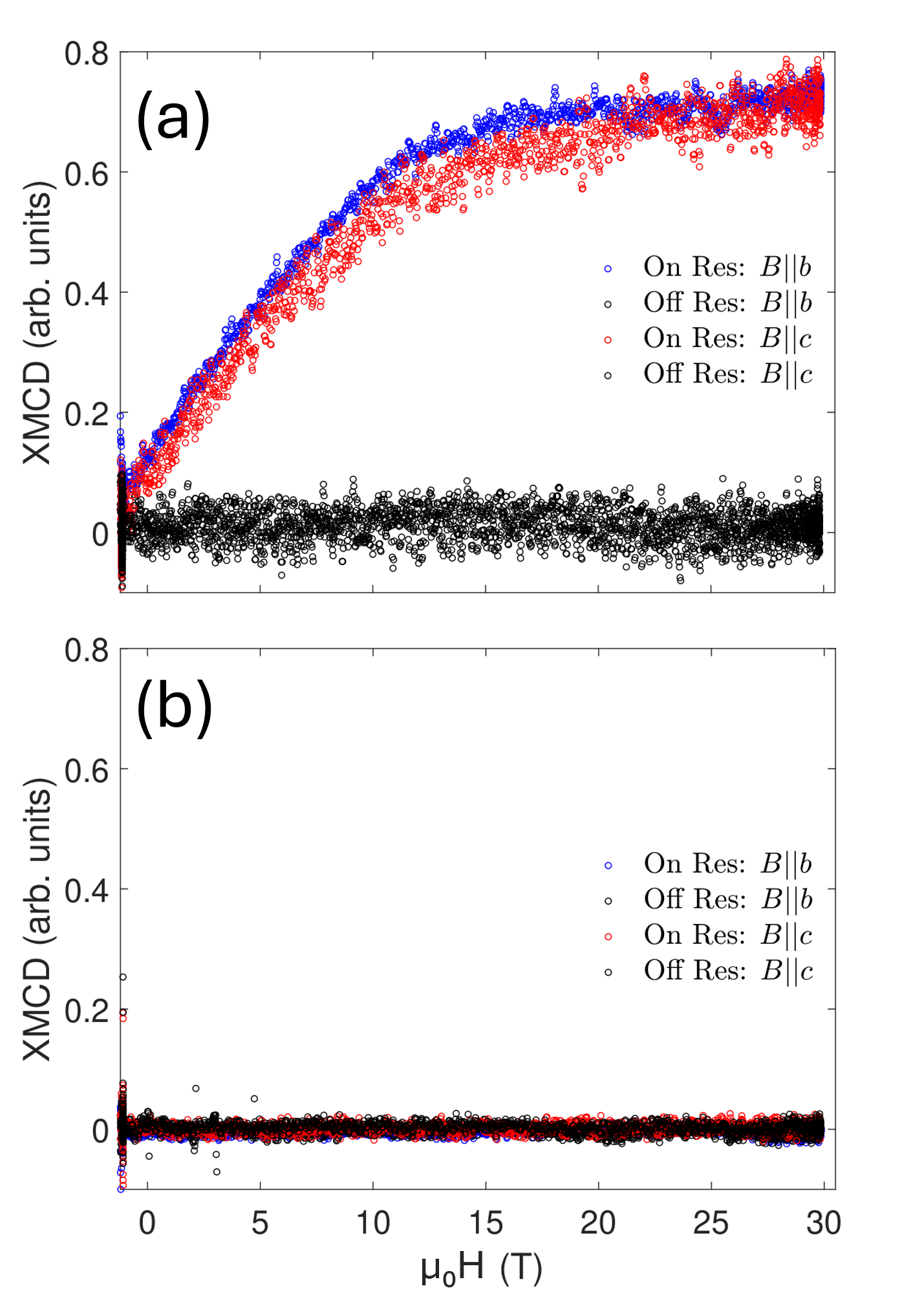}
    \caption{(a) XMCD measured at the Eu $M_5$ edge (1130\,eV) at 10\,K with pulsed field applied along the $b$ and $c$ axes. (b) Mn $L_3$ edge (640.5\,eV) XMCD data along the same directions and same temperature. The off-resonance energies are 1125\,eV and 635\,eV respectively.}
    \label{fig:xmcd}
\end{figure}

Whether the Mn magnetic sublattice changes below $T_{\text{Eu1}}$ and in applied field is important for an accurate refinement for the Eu sublattice spin structure. To resolve this question, We performed pulsed field XMCD to detect any change at the Mn and Eu absorption edges. The data presented in Fig~\ref{fig:xmcd} shows the ferromagnetic component of the spin sublattices of Eu and Mn as a function of applied field at $T = 10$\,K, which is below $T_{\text{Mn}}$ and $T_{\text{Eu1}}$ but above $T_{\text{Eu2}}$. Fig~\ref{fig:xmcd}(a) shows XMCD data measured at Eu $M_5$ edge while (b) shows data measured at Mn $L_3$ edge. As XMCD is particularly sensitive to the ferromagnetic component of the spins along the x-ray
propagation direction (= field direction), it indicates how far the spins cant away from the  AFM structure in zero field. The Eu $M_5$ edge XMCD data shows that the Eu spins start canting at a very low field and starts to saturate at around 15\,T for both $H \parallel b$ and $H \parallel c$. This is in agreement with our pulsed field data of the magnetisation (see below). The latter reveal a cusp-like feature at around 0.5\,T, which is prominent below 10\,K, and cannot be observed in the pulsed-field XMCD data. This feature will be discussed in detail below. On the other hand, one can see that the XMCD at Mn $L_3$ edge remains absent up to 30\,T. This supports the assumption that the exchange coupling between Mn spins is strong enough that the Mn spin structure remains collinear without any field-induced canting at low temperatures, at least up to 30\,T. 

\begin{figure*}
    \centering
    \includegraphics[width=\textwidth]{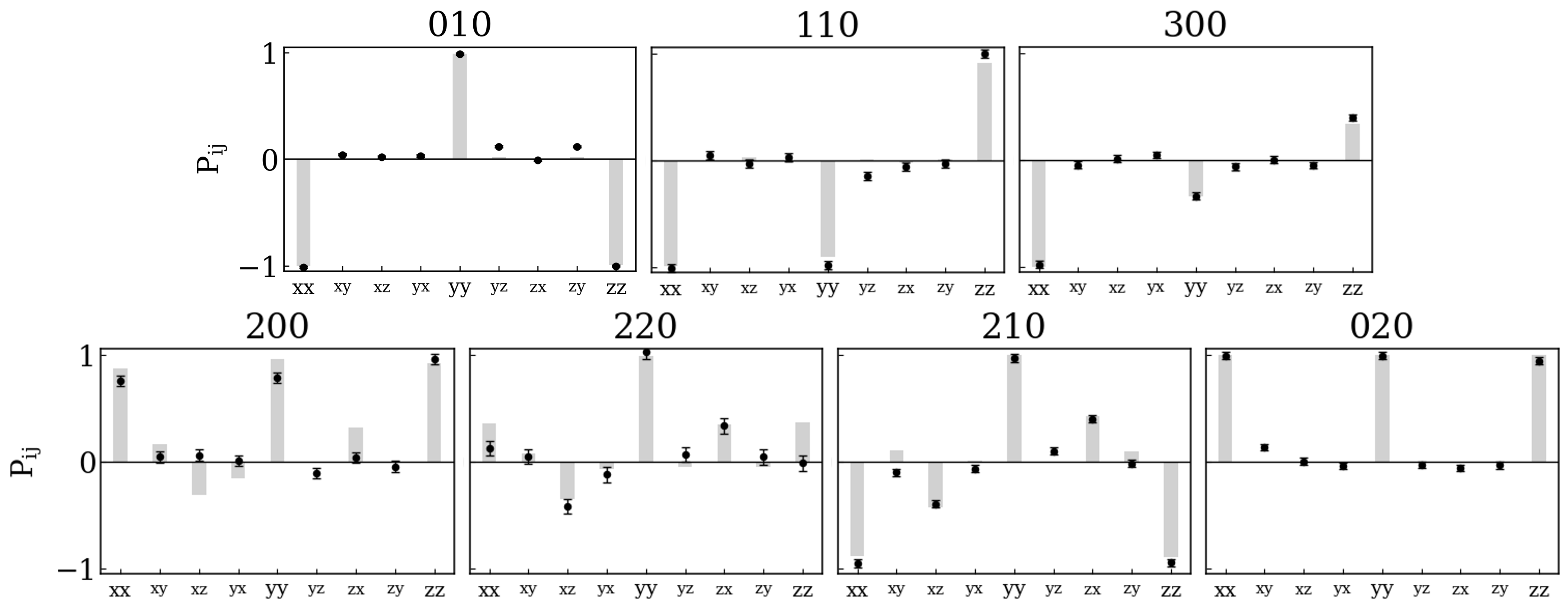}
    \caption{Polarisation matrices measured by SNP at a temperature of 7.5\,K. The upper row are pure magnetic Bragg peaks. The (200), (220) and (210) peaks in the lower row have contributions from nuclear magnetic interference (off-diagonal terms), and the (020) peak is purely structural. The grey bars are calculated from the magnetic structure shown in Fig.~\ref{fig:3} combined with the crystal structure.}
    \label{fig:2}
\end{figure*}

\begin{figure}
    \centering
    \includegraphics[width=0.6 \columnwidth]{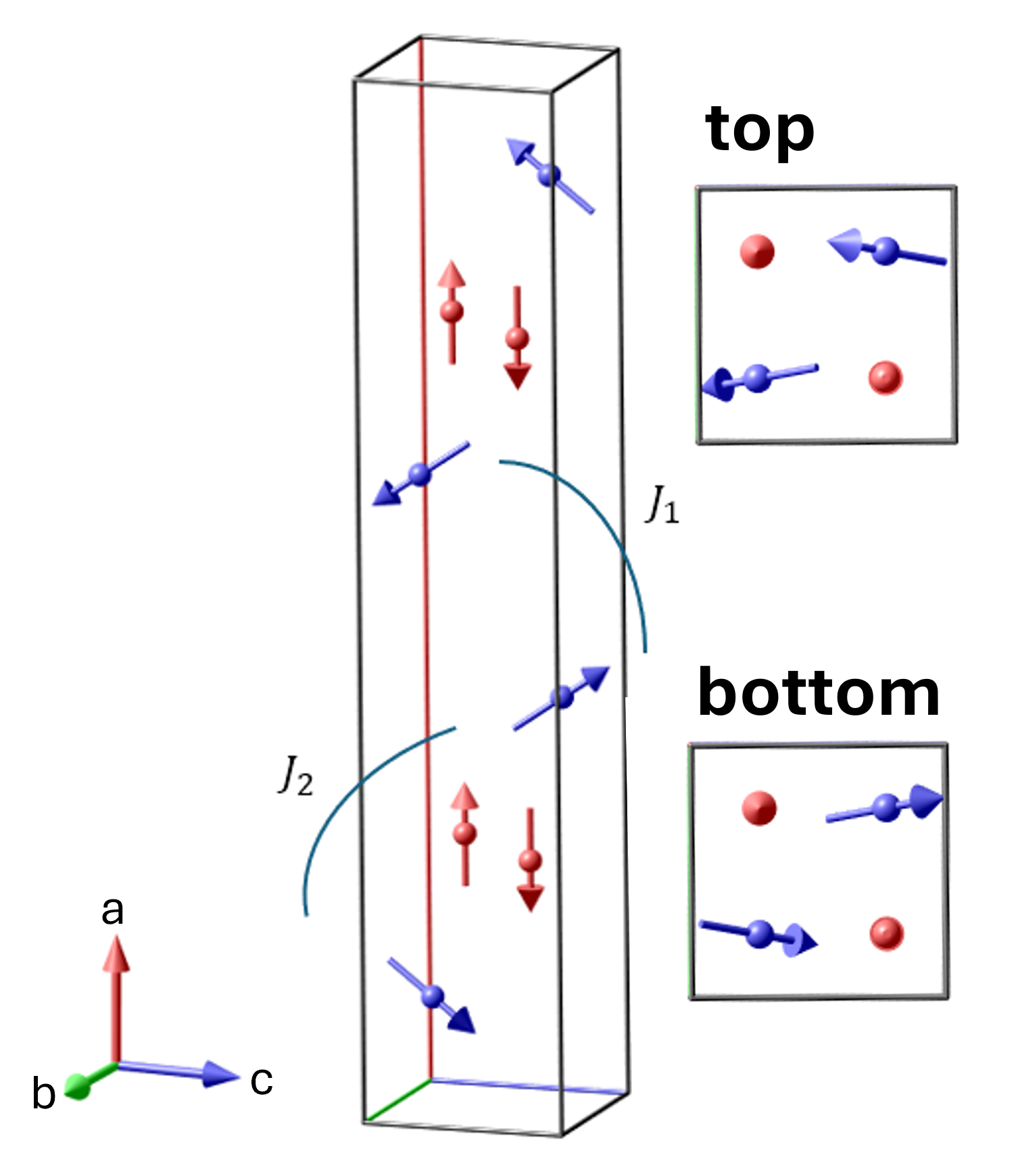}
    \caption{Refined magnetic structure for $T = 2$\,K $<T_{\text{Eu2}}$ showing the AFM and FM exchanges $J_{1}$,$J_{2}$ between the Eu magnetic moments. The top and bottom halves of the unit cell are shown separately in the inset to show the small canting along $b$ axis.}
    \label{fig:3}
\end{figure}

Moving on to the Eu magnetic sublattice, we represent in Fig.~\ref{fig:2} the polarization matrices measured at several reflections at 7.5\,K. This temperature is slightly below  $T_{\text{Eu2}} \simeq 9$\,K for  our Sn-flux sample (Fig.~\ref{fig:4}). After refining the spin structure based on our SNP measurements and the assumption (supported by the XMCD data) that Mn spins remain in the C-type AFM structure we find that the (010) peak is purely magnetic and derives solely from Mn AFM structure,  the  (110) and (300) peaks are purely Eu magnetic peaks, and the (200), (220), (210) peaks have both nuclear and magnetic contributions. The (020) is a reference structural Bragg peak.  The grey bars in Fig.~\ref{fig:2} show the calculated $P_{ij}$ for the refined magnetic structure, in which neighboring Eu spins that lie between the Mn layers remain antiparallel but are canted away from both the $a$ and $c$ axes as shown in Fig~\ref{fig:3}. The non-zero off-diagonal components in the matrices for peaks with nuclear--magnetic interference, such as (210) and (220), continue to imply a significant domain imbalance at this temperature.

The refined canting angle in the $ac$ plane is approximately 35$^\circ$ away from the $a$ axis, while the canting angle in the $bc$ plane is $\sim$ 15$^\circ$ from $c$. The $ac$ canting angle is quite close to previous reports of 31$^\circ$ at 12\,K~\cite{Wilde} and 41$^\circ$ at 7\,K~\cite{Gong}, but the $bc$ canting angle is much less than the $\sim$ 50$^\circ$ reported in Ref.~\onlinecite{Wilde} at 5\,K. Measurements were also made at 2\,K to check the structure closer to the ground state. Small  increases in the refined canting angles were obtained, with the 2\,K values being 38$^\circ$ and 16$^\circ$ in the $ac$ and $bc$ planes, respectively. 

\begin{figure}
    \centering
    \includegraphics[width=0.7 \columnwidth]{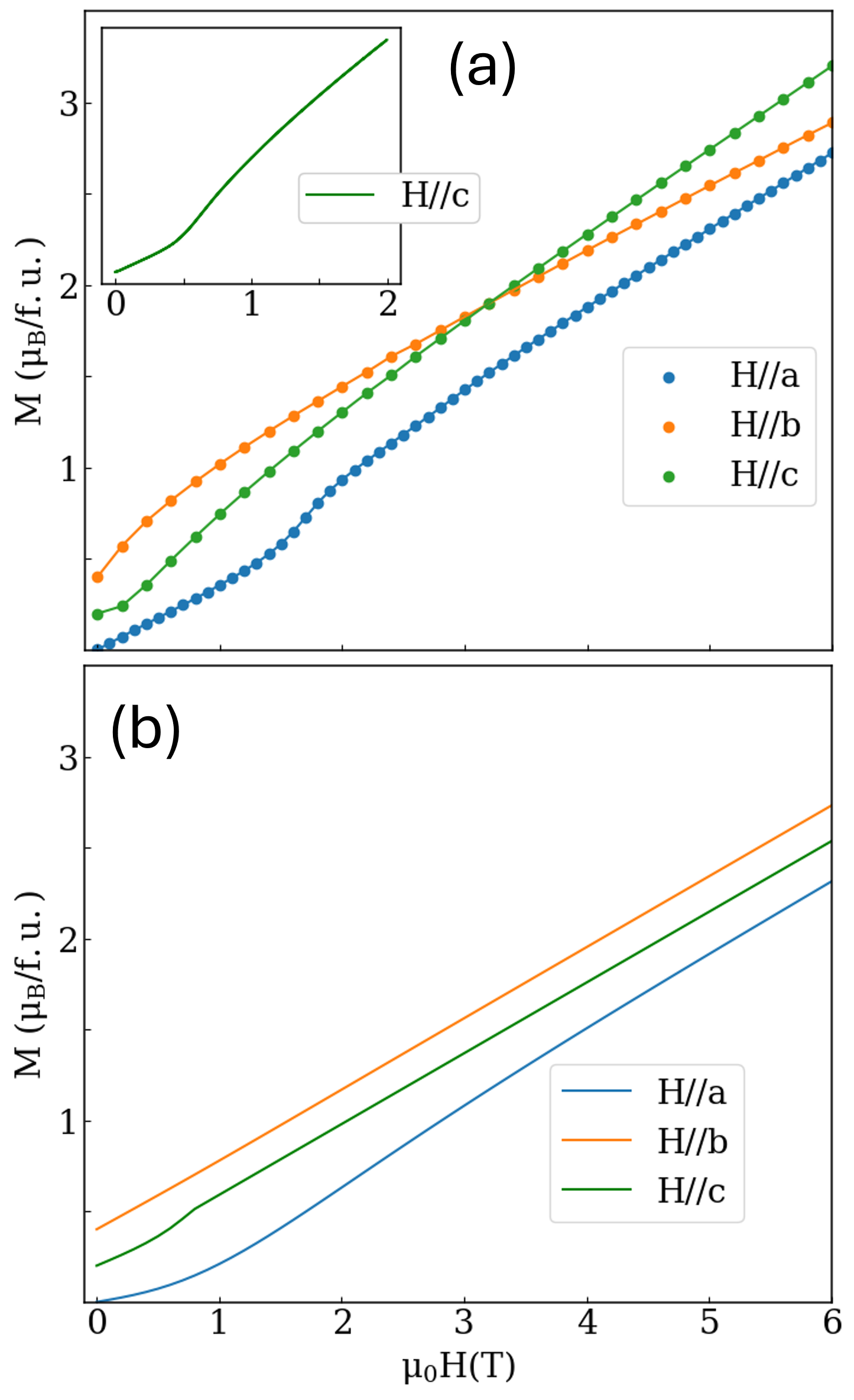}
    \caption{(a) Magnetization data for field along the $a$, $b$ and $c$ axes and $T = 2$\,K. The curves have been offset vertically for clarity. Spin-flop transitions can be observed at $\mu_0 H =0.5$\,T ($H \parallel c$) and 1.5\,T ($H \parallel a$). The inset shows the low field part of the $H \parallel c$ data. (b) Simulated magnetization curves from the mean-field model (see text).}
    \label{fig:6}
\end{figure}

\begin{figure}
    \centering
    \includegraphics[width=0.7 \columnwidth]{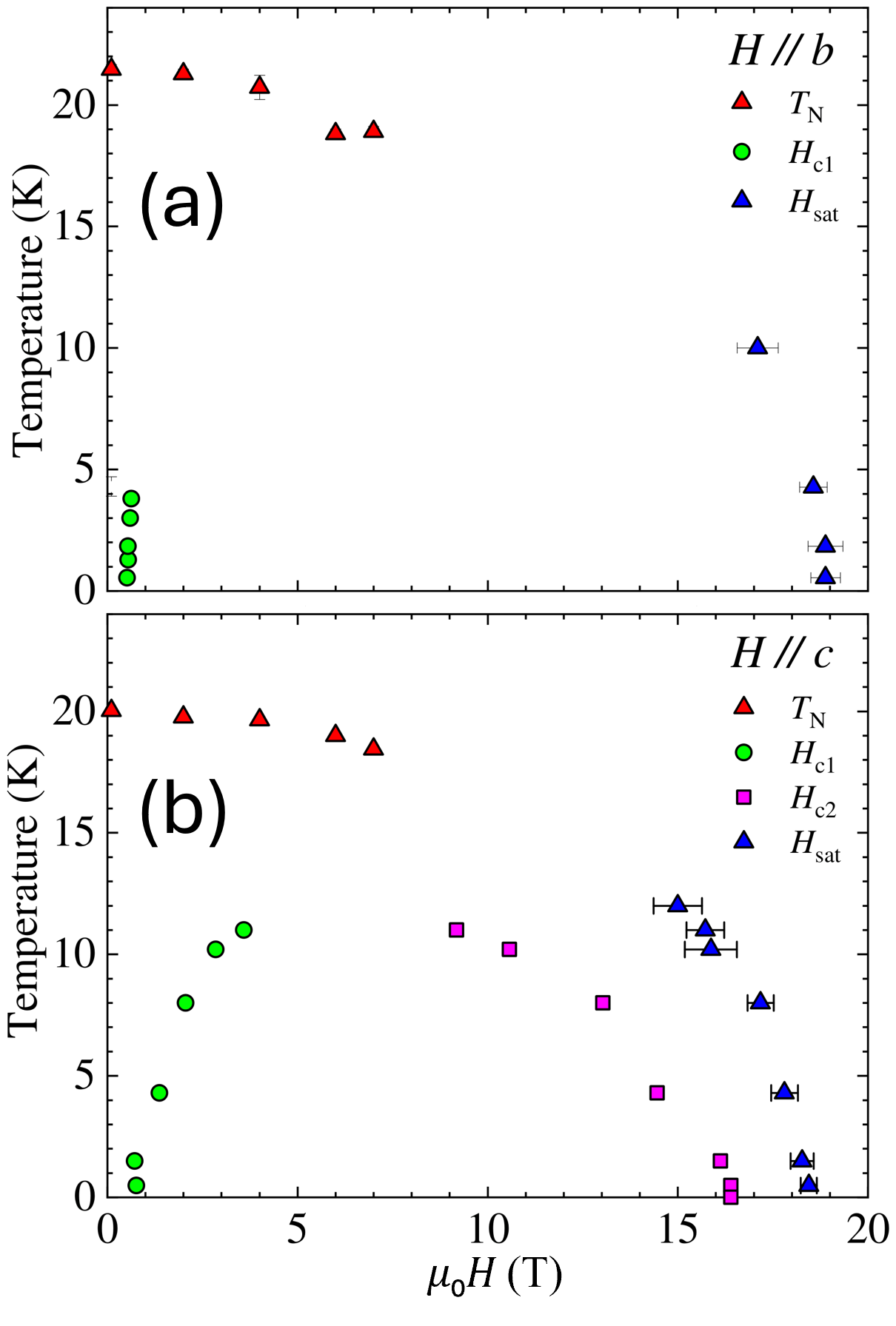}
    \caption{Phase diagram for (a) $H\parallel b$ and (b) $H\parallel c$, showing critical fields extracted from isothermal pulsed-field magnetisation and Eu ordering temperatures from low-field temperature dependent susceptibility measurements. Only $H\parallel c$ measurements show the additional peak at $H_{c2}$ before the magnetisation saturates. Error bars shown where larger than the point size.}
    \label{fig:10}
\end{figure}

Figure~\ref{fig:6}(a) shows magnetization measurements up to 6\,T and $T = 2$K, with $H$ along all three crystallographic axes.  For $H \parallel a$, a clear transition can be seen around 1.5\,T. This has previously been interpreted as a spin-flop transition~\cite{Yi,Gong,Sun}. The inset shows an enlargement of the $H \parallel c$ curve below $\mu_0 H =2$\,T. One can observe a clear spin-flop transition in this direction around 0.5\,T. Finally, the $H \parallel b$ curve shows no such transition and in fact has a negative curvature as field increases. A mean-field model which attempts to explain the behavior of the magnetization in terms of a microscopic Hamiltonian is developed in the next section. The change of gradient at $T_{\rm Eu1}$ moves to lower temperatures as the applied field is increased. The values obtained for $H \parallel b$ and $c$ at different fields are shown in the phase diagrams of Fig.~\ref{fig:10}.

\begin{figure}
    \centering
    \includegraphics[width=0.9 \columnwidth]{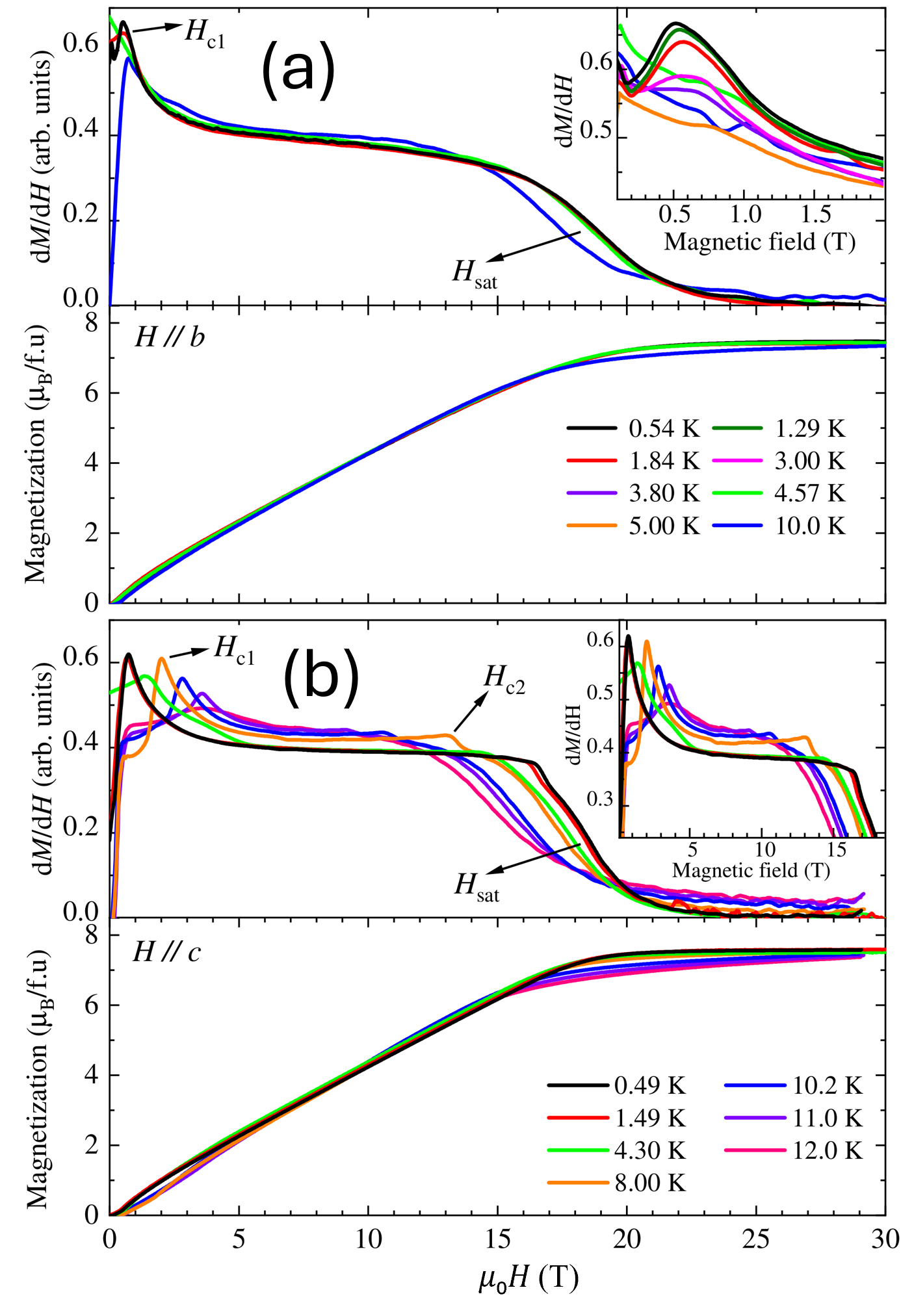}
    \caption{Pulsed-field magnetization with field along (a) the $b$ axis, and (b) the $c$ axis. The top panels show $\textrm{d}M/\textrm{d}H$. The peaks in $\textrm{d}M/\textrm{d}H$ at low field marked $H_{\textrm{c}1}$ are due to spin-flop transitions. The magnetization starts to saturate around 15\,T and fully saturates around 19\,T, depending on temperature. In (b), a second anomaly, marked $H_{\textrm{c}2}$, can be seen in $\textrm{d}M/\textrm{d}H$ just below the saturation field }
    \label{fig:5}
\end{figure}

Before that, we discuss the pulsed-field magnetization measurements shown in Fig.~\ref{fig:5}. The magnetization exhibits some low-field anomalies followed by a roughly linear increase with field before eventually saturating at a field of approximately 19\,T at the lowest temperature. To study the subtle changes of magnetization at high field, we plot $\textrm{d}M/\textrm{d}H$ in the top panels of Fig.~\ref{fig:5}(a) and (b) for fields along the $b$ and $c$ axis, respectively. For both directions, there is a peak denoted $H_{\textrm{c}1}$ that corresponds to the spin-flop transition, and the magnetization saturates at $H_{\text{sat}}$. However, for $H \parallel c$ we also observe an additional step, denoted $H_{\textrm{c}2}$, which occurs a few tesla below $H_{\text{sat}}$.  A summary of the critical fields for both orientations at different temperatures is presented in Fig.~\ref{fig:10}, along with the $T_{\rm Eu1}$ extracted from the temperature dependence of the susceptibility. The $H_{\textrm{c}2}$ feature suggests an additional field-induced spin-reorientation transition when $H \parallel c$. We will discuss the simulation of this spin-reorientation transition in our mean-field model in the following section.

\section{Mean-field Model}

We first consider the ground state magnetic structure with zero field. To start, we consider the nearest-neighbor AFM exchange, denoted $J_{1}$, for Eu moments between Mn layers, and the FM exchange $J_{2}$ for neighboring moments separated by Mn layers (Fig.~\ref{fig:3}). These exchanges alone are not enough to produce the canted structure. Attempts were made to add other terms such as single-ion anisotropy, but the parent symmetry of the crystal structure would cancel out effects of such anisotropy. 

Instead, we consider the exchange coupling between the Mn AFM sublattice and the Eu moments. Since the Mn moments are assumed to be rigid in keeping with the XMCD results, they are fixed along the $a$ axis regardless of applied field, acting as an effective staggered field on the Eu moments. Inclusion of this staggered field along  $a$ allows a ground state with canting in the $ac$ plane to be stabilized. 

To introduce the canting towards the $b$ axis, we add another staggered field in a similar fashion, this time along $b$. The full effective spin Hamiltonian is described below: 

\begin{align}
\mathcal{H}
&= - J_{1} \sum_{\langle i,j\rangle} 
      \mathbf{S}^{\mathrm{Eu}}_{i}\!\cdot\!\mathbf{S}^{\mathrm{Eu}}_{j}
   - J_{2} \sum_{\langle\!\langle i,j\rangle\!\rangle} 
      \mathbf{S}^{\mathrm{Eu}}_{i}\!\cdot\!\mathbf{S}^{\mathrm{Eu}}_{j}
   \nonumber \\[4pt]
&\quad
   - g_{\mathrm{Eu}}\mu_{\mathrm{B}}\sum_{i} \mathbf{B}\!\cdot\!\mathbf{S}^{\mathrm{Eu}}_{i}
   - \sum_{i} \big( h_{a}\,\sigma_{i}\,\hat{\mathbf{a}} + h_{b}\,\tau_{i}\,\hat{\mathbf{b}} \big)
                \!\cdot\!\hat{\mathbf{S}}^{\mathrm{Eu}}_{i}\, .
\label{eq:H_two_stag}
\end{align}

\begin{equation}
\sigma_{i},\  \tau_{i}=\begin{cases}+1,& i\in A \\-1,& i\in B\end{cases}
\label{eq:stag}
\end{equation}
where $A,B$ denotes neighboring layers of Eu sublattices. $\sigma_{i},\tau_{i}$ are alternating factors taking account of Eu moments between and across the Mn layers to represent the direction of the staggered fields. $h_{a}$, $h_{b}$ are amplitudes of the staggered field, and their ratio with the exchange interaction strengths $J_{1}, J_{2}$ can be determined by the critical field of the spin flop transitions along with the refined canting angles. In addition, the value and ratio of the exchange interactions were determined from saturation field of the magnetization given in Fig~\ref{fig:5}. The values of the Hamiltonian parameters are listed in Table~\ref{table}

\begin{table}[h]
\centering

\begin{tabular}{c c}
\hline
Parameter & Value (meV) \\
\hline
$J_1$ &  \hspace{-7pt}$-0.149$\\
$J_2$ &  0.002\\
$h_a$ &  0.081\\
$h_b$ &  0.021\\
\hline
\end{tabular}
\caption{Model parameters used in the mean-field simulation.}
\label{table}
\end{table}

\begin{figure*}
    \centering
    \includegraphics[width=\textwidth]{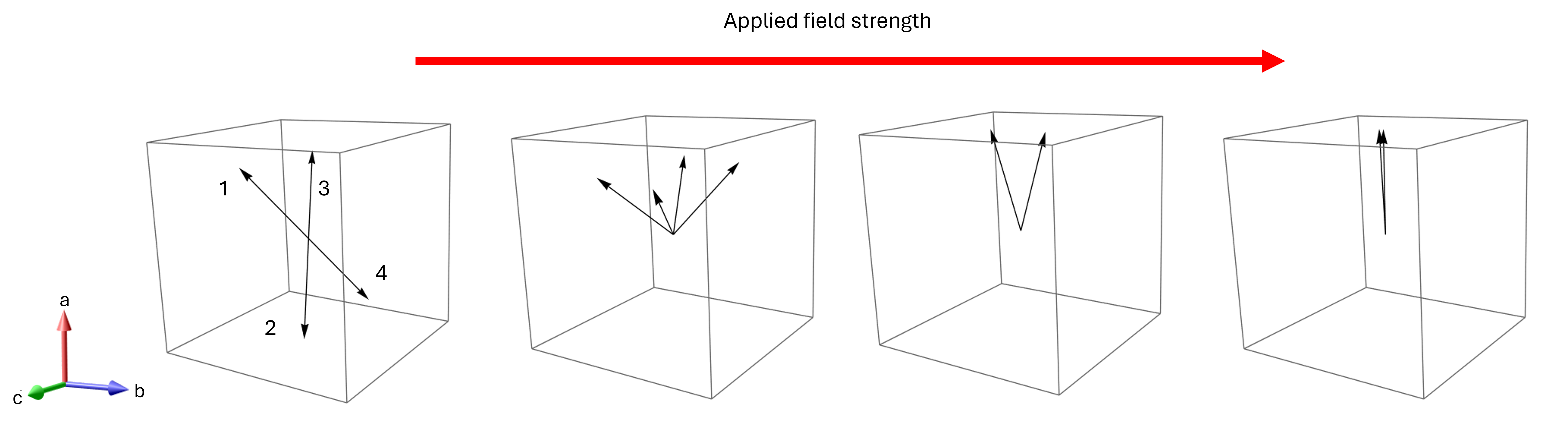}
    \caption{Evolution of Eu spins with increasing  field applied parallel to the $a$ axis. Spins 1 to 4 are labeled from top to bottom in the unit cell of Fig.~\ref{fig:3} (From left to right: $\mu_0 H = 0$\,T, 8\,T, 16\,T, 20\,T).}
    \label{fig:spin dynamics}
\end{figure*}

The staggered-field terms in the Hamiltonian are essential for reproducing the experimentally determined canted ground state. They phenomenologically represent a net effective field exerted by the ordered Mn sublattice on the Eu moments. Since the Mn moments are oriented predominantly along the $a$ axis, the longitudinal component $h_a$ is naturally much larger than $h_b$. The smaller $b$-axis component cannot be generated by a purely isotropic Heisenberg interaction with the Mn moments. Instead, it requires an anisotropic component of the Eu--Mn exchange tensor that couples an $a$-axis Mn moment to a $b$-axis Eu moment. 
The symmetry of these paths causes the effective $b$-axis field to alternate between neighbouring Eu layers, thereby producing the observed canting pattern. We emphasize that $h_a$ and $h_b$ are phenomenological parameters in the present model; establishing their precise microscopic origin would require a detailed calculation of the Eu--Mn exchange tensor.

We now turn our attention to the field-induced spin flop transitions. The zero-temperature simulated magnetization curves are shown in Fig~\ref{fig:6}(b). The model successfully captures the spin-flop transition for $H\parallel c$ at $\mu_0 H = 0.5$\,T with a cusp in the magnetization curve. With increasing field, the Eu spins which have $c$-axis component opposite to the field when $H < H_{\textrm{c}1}$  rotate so that the $c$-axis component is along the field when $H > H_{\textrm{c}1}$, while the other Eu spins move only slightly on crossing $H_{\textrm{c}1}$. For $H \parallel a$, the experimental magnetization shows a linear increase up to the spin-flop transition at 1.5\,T, Fig~\ref{fig:6}(a), and the model gives a similar albeit smoother curve. For $H \parallel b$ there is a significant discrepancy between simulation and experiment: the model predicts a linear relationship for the whole field range but the data show a clear negative curvature. Overall, the simple mean-field model reproduces the $H_{\textrm{c}1}$ field-induced transitions  observed for $H \parallel a$ and $H \parallel c$ reasonably well.

\begin{figure}
    \centering
    \includegraphics[width=0.7 \columnwidth]{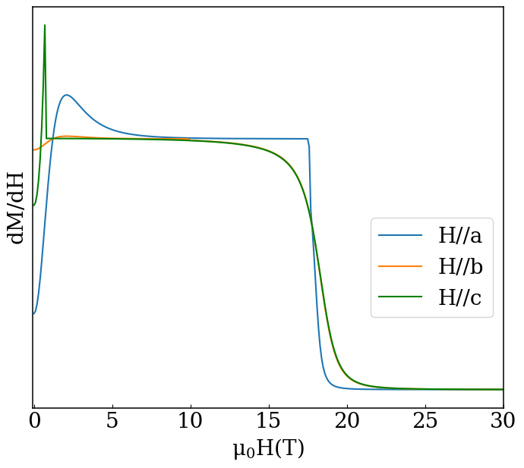}
    \caption{Simulated $\textrm{d}M/\textrm{d}H$ curves as a function of field from the mean-field model. For $H \parallel a$, the step-like drop in $\textrm{d}M/\textrm{d}H$ at $\mu_0H = 15.5$\,T corresponds to the alignment of pairs of spins shown in Fig.~\ref{fig:spin dynamics} and is reminiscent of the $H_{c2}$ feature in the $H \parallel c$ data. }
    \label{fig:dM/dH}
\end{figure}

The field-induced transition at $H_{\textrm{c}2}$ observed for $H \parallel c$ is more puzzling. According to the model, when $H \parallel c$ the Eu spins rotate continuously towards the $c$ axis with increasing field. No detectable $H_{\textrm{c}2}$ anomaly is found in the simulated curve. On the other hand, when $H \parallel a$ an $H_{\textrm{c}2}$-like feature is observed in the simulation. The origin of this transition is represented in Fig.~\ref{fig:spin dynamics}, in which the Eu spins are labeled 1 to 4 from top to bottom in the unit cell shown in Fig.~\ref{fig:3}. One can see that just before saturation the four spins form into two aligned pairs. This happens at the step-like feature in the simulation shown in Fig.~\ref{fig:dM/dH}, which resembles the experimentally observed $H_{\textrm{c}2}$ feature. It is likely, therefore, that the $H_{\textrm{c}2}$ for $H \parallel c$ corresponds to a similar spin alignment just before saturation.

\section{Conclusion \& Outlook}
We have investigated the magnetic ground state and field-induced magnetic transitions of the magnetic Dirac semimetal candidate EuMnSb$_{2}$ using spherical neutron polarimetry, unpolarized neutron diffraction, pulsed-field XMCD, and magnetization measurements. By performing these measurements on the same Sn-flux-grown single crystal, we avoid ambiguities associated with sample dependence and provide a consistent description of the Eu and Mn magnetic sublattices. The magnetic order was determined conclusively by SNP to avoid the neutron absorption problem. The Mn sublattice is confirmed to adopt the previously reported C-type antiferromagnetic structure with moments along the $a$ axis. Below the Eu ordering temperature, the obtained Eu spin structure is essentially the same as that found by Wilde \textit{et al.}~\cite{Wilde} but with a smaller $b$ axis component of the Eu spins. 

Field-dependent magnetization measurements reveal low-field spin-flop transitions for fields applied along the $a$ and $c$ axes, together with a higher-field anomaly $H_{c2}$ for $H\parallel c$ before saturation. Pulsed-field XMCD measurements at the Eu $M$ and Mn $L$ edges show that these field-induced transitions are associated primarily with the Eu sublattice, while the low-temperature Mn magnetic structure remains unchanged within the sensitivity of the experiment up to the highest fields measured. This establishes the Mn sublattice as an effectively rigid magnetic background over the investigated field and temperature range.

To rationalize these observations, a minimal mean-field model has been developed that describes the ground-state canted structure. The model includes effective staggered fields to represent the Eu--Mn coupling. The  magnetic interactions are found to be dominated by the AFM coupling $J_1$, and the observed spin-flop transitions for $H\parallel a$ and $H\parallel c$ are qualitatively captured. On the other hand, the model does not reproduce the $H \parallel b$ magnetization particularly well, nor does it account for the $H_{\textrm{c}2}$ anomaly observed for $H\parallel c$. It is possible that these shortcomings might signify the need for additional terms in the Hamiltonian.

One possible microscopic origin of the effective staggered coupling is an antisymmetric exchange interaction between the Eu and Mn sublattices. In particular, the absence of inversion symmetry at relevant Eu–Mn exchange paths allows Dzyaloshinskii–Moriya interactions by symmetry, which could generate transverse effective fields on the Eu moments in the presence of the ordered Mn background. The dipolar interaction could be another possible source of the anisotropic coupling between Mn and Eu spins. Such interactions provide a natural route to stabilizing the observed canting of the Eu spins, although a quantitative microscopic model of these interaction pathways remains to be developed.

Several questions therefore remain open: (1) Why does such a large magnetic domain imbalance occur, as evidenced from the non-zero off-diagonal elements of $P_{ij}$ for the Bragg peaks (210) and (220)? (2) What is the origin of the signal denoted $H_{\textrm{c}2}$ observed in the high-field magnetization for $H\parallel c$? Is it related to the flop-like transition predicted by the model for $H \parallel a$?  (3) What is the microscopic mechanism that couples the Eu and Mn spins and drives the canted structure?

By establishing a conclusive result for the magnetic structure and modeling its magnetic interactions, our work has proposed a framework to understand the magnetic structure of $A$Mn$X_{2}$-type compounds in an applied field. The results may provide insights into how the topology can be tuned with magnetic field in this class of compounds.

\acknowledgements

We acknowledge Ivica Zivkovic for help with magnetization measurements and sample preparation, Jennifer Sears and Sonia Francoual for help with the preliminary REXS study at P09, DESY. We acknowledge support from the UK Research and Innovation (UKRI) under the UK government’s Horizon Europe funding guarantee [Grant No. EP/X025861/1]. The pulsed magnet system as the Nicholas Kurti High Magnetic Field Laboratory was refurbished with a grant from the UK
Engineering and Physical Sciences Research Council
(EPSRC) [Grant No. EP/J013501/1]. This work is based on experiments performed on ZEBRA at the Swiss spallation neutron source SINQ, Paul Scherrer Institute, Villigen, Switzerland. The proposal numbers for the data presented in this manuscript are 5-55-21~\cite{D3_data} (D3, ILL) and 5-41-1286~\cite{D9_data} (D9, ILL). Y.F.C. acknowledges funding from the Diamond Light Source and the Clarendon Scholarship from the University of Oxford under joint doctoral studentship no. STU0477. S.S. thanks the EPSRC for funding. P.A.G. thanks Ewan Jewkes and Patrick Ruddy for technical assistance. The National High Magnetic Field Laboratory is supported by National Science Foundation Cooperative Agreement No. DMR-2128556 and the Department of Energy (DOE). J.S. is grateful to the University of Oxford for the provision of a visiting professorship that permitted some of the work described in this paper. A.T.B. and D.P. acknowledge support from the Oxford–ShanghaiTech collaboration project.

\bibliography{biblio}

\end{document}